\documentclass[a4paper]{jpconf}
\usepackage{graphicx}
\usepackage{cite}
\begin{document}
\title{CYGNO: Triple-GEM Optical Readout for Directional Dark Matter Search}

\author{
 I.~Abritta~Costa$^1$,
 E.~Baracchini$^2$,
 R.~Bedogni$^3$,
 F.~Bellini$^4$,
 L.~Benussi$^3$,
 S.~Bianco$^3$,
 M.~Caponero$^{3,7}$,
 G.~Cavoto$^4$,
 E.~Di Marco$^1$,
 G.~D'Imperio$^1$,
 G.~Maccarrone$^3$,
 M.~Marafini$^{1,5}$, 
 G.~Mazzitelli$^3$,
 A.~Messina$^4$,
 F.~Petrucci$^6$,
 D.~Piccolo$^3$,
 D.~Pinci$^{1,}\footnote[9]{Corresponding author}$,
 F.~Renga$^1$,
 G.~Saviano$^{3,8}$ and
 S.~Tomassini$^3$}

\address{
$^1$INFN - Sezione di Roma \\
$^2$Gran Sasso Science Institute L'Aquila \\
$^3$INFN - Laboratori Nazionali di Frascati \\
$^4$Dipartimento di Fisica - Sapienza Universit\`a di Roma \\
$^5$Museo Storico della Fisica e Centro Studi e Ricerche "Enrico Fermi" \\
$^6$INFN - Sezione di Roma TRE \\
$^7$ENEA - Frascati \\
$^8$Dipartimento di Ingegneria Chimica, Materiali e Ambiente - Sapienza Universit\`a di Roma}

\ead{davide.pinci@roma1.infn.it}

\begin{abstract}
CYGNO is a project realising a cubic meter demonstrator to study the scalability of the performance of the optical approach for the readout of large-volume, GEM-equipped TPC. This is part of the CYGNUS proto-collaboration which aims at constructing a network of underground observatories for directional Dark Matter search.
The combined use of high-granularity sCMOS and fast sensors for reading out the light produced in GEM channels during the multiplication processes was shown to
allow on one hand to reconstruct 3D direction of the tracks, offering accurate energy measurements and sensitivity to the source directionality and, on the other hand, a high particle identification capability very useful to distinguish nuclear recoils.
Results of the performed R\&D and future steps toward a 30-100 cubic meter experiment will be presented.
\end{abstract}

\section{Introduction}
This paper summarises the work performed by the CYGNO collaboration in 
developing a Time Projection Chamber with an electron 
multiplication stage based on triple-GEM with an optical readout.

This activity is carried out in the framework of the CYGNUS-TPC project \cite{bib:0} that aims at building a multi-ton gas target distributed in underground labs using TPC techniques or Directional Dark Matter search.

Time projection chambers provide very useful information:
\begin{itemize}
    \item it is possible to make a 3D reconstruction of the tracks in sensitive volume;
    \item it can evaluate not only the total amount of released energy, but also its profile along the particle trajectory allowing to reconstruct the dE/dx, very helpful for particle identification and track head-tail discrimination;
    \item large volumes can be acquired with small amount of readout channels. 
\end{itemize}

For equipping large surfaces, the use of Micro Pattern Gas Detectors
is a very simple solution ensuring high space and time resolution.
In particular Gas Electron Multipliers are able to suppress
the Ion Back Flow inside the sensitive volume. 

During the multiplication processes, 
photons are produced along with electrons 
by the gas through atomic and molecular de-excitation.

Optical readout of the light produced by
GEM was already studied in the past 
\cite{bib:ref1, bib:ref2, bib:ref3, bib:ref4, bib:ref5}
and showed several advantages:
\begin{itemize}
\item optical sensors offer higher granularities with
respect to electron sensitive devices;
\item optical coupling allows to keep sensor out of the 
sensitive volume reducing the interference with high voltage operation 
and lowering the gas contamination;
\item the use of suitable lens allows to 
acquire large surfaces with small sensors;
\end{itemize}

A gas represents an interesting target: nuclei free paths can be long enough to be reconstructed. 
In particular, $\epsilon$ the maximum fraction of the energy that can be
transferred to the nucleus of mass $m_N$ by a Dark Matter particle of mass
$m_{\chi}$ is given by:
\begin{equation}
\epsilon = \frac {4 \rho}{\left( \rho + 1 \right)^2}
\label{eq:trans}
\end{equation}

where $\rho = \frac{m_N}{m_{\chi}}$.

\subsection{LEMON detector}

\begin{figure}[htbp]
\centering
\includegraphics[width=.85\textwidth]{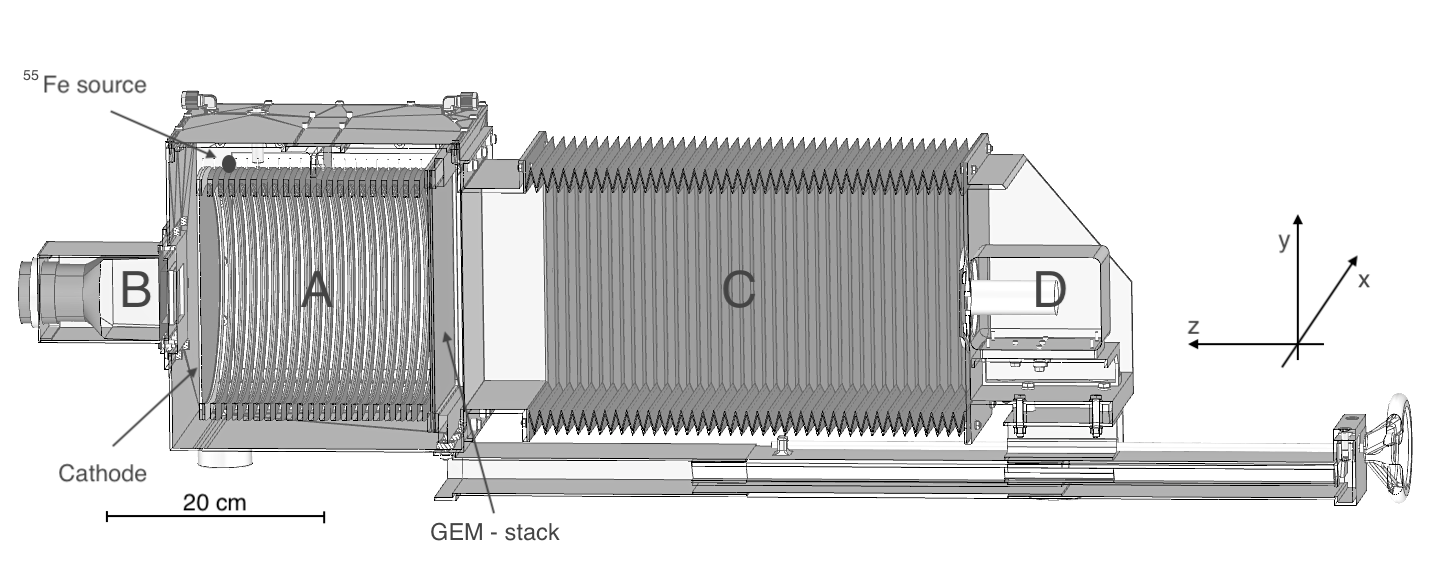}
\caption{Drawing of the experimental setup.
In particular, the elliptical field cage close on one side by the triple-GEM structure and on the other 
side by the semitransparent cathode (A), 
the PMT (B), 
the adaptable bellow (C) and the CMOS camera with its lens (D) 
are visible.}
\label{fig:lemon}
\end{figure}

In order to test the performance of the optical approach, several prototypes were tested, and, in particular, all measurements described in this work were carried out on the Large Elliptical MOdule (LEMON, Fig. \ref{fig:lemon})
which is described in details in refs. \cite{bib:eps, bib:ieee17, bib:ieee18}.

The baseline gas mixture is an He/CF$_4$ 60/40 that showed to provide very clear optical signals \cite{bib:nim_orange1, bib:nim_orange2, bib:jinst_orange2}.
According to eq. \ref{eq:trans}, the  use of He allows an efficient momentum transfer from Dark Matter (DM) particles of 1 GeV mass ($\epsilon_{He} \simeq$ 60\%, $\epsilon_{Ar} \simeq$ 10\% and $\epsilon_{Xe} \simeq$ 3\%).

\section{Operative threshold evaluation}

To get an indication of the reliable detection threshold, the sensor noise level was studied and compared to the detector response to 5.9 keV photons provided by a $^{55}$Fe source.

\subsection{Measurements with a $^{55}$Fe source}

\begin{figure}[htbp]
\centering
\includegraphics[width=.45\textwidth]{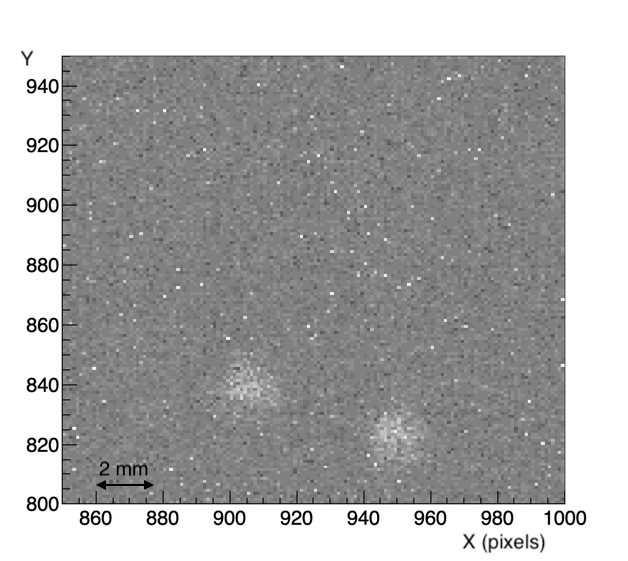}
\caption{Example of two clusters due to X-ray interaction in gas.}
\label{fig:spot}
\end{figure}

Figure~\ref{fig:spot} shows an example of an image of two light 
spots due the interaction of the $^{55}$Fe photons in the gas.

\begin{figure}[htbp]
\centering
\includegraphics[width=.25\textwidth, angle=270]{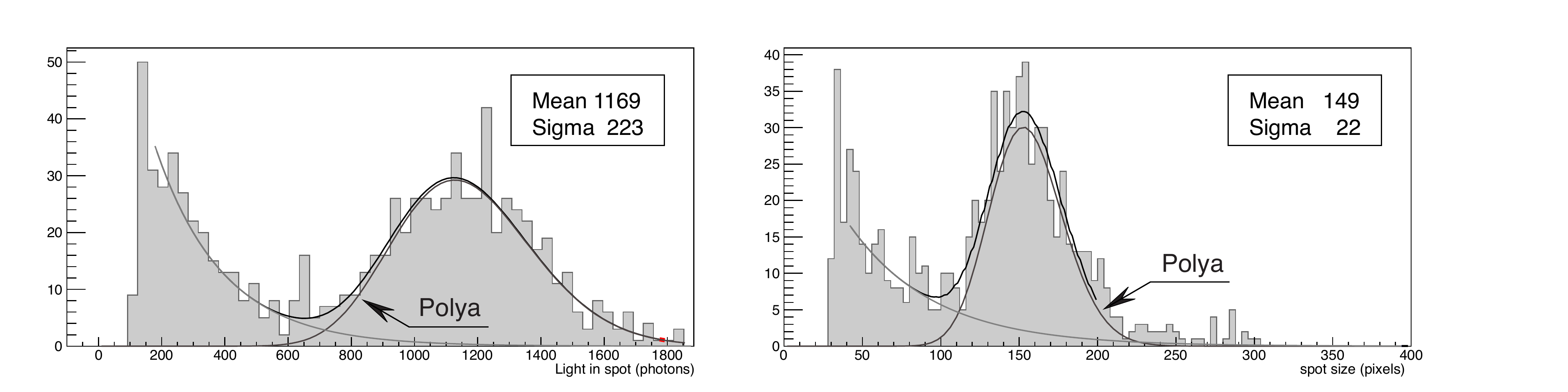}
\caption{Distribution of total light (left) and number of illuminated pixels (right) for a run taken with V$_{\rm GEM}$~=~450 V, E$_{\rm d}$~=~600 V/cm and E$_{\rm t}$~=~2 kV/cm.}
\label{fig:spectra}
\end{figure}

The spectrum of the total light in clusters reconstructed within the sensitive area and the distribution of their size (i.e. the number of over-threshold pixels) are studied.
Figure \ref{fig:spectra} shows an example of these distributions for a run taken with V$_{\rm GEM}$~=~450 V, E$_{\rm d}$ = 600~V/cm and E$_{\rm t}$ = 2~kV/cm.

In order not to slow down the analysis procedure, only clusters with at least 30 pixels over-threshold were considered. The distribution is fitted with the sum of an exponential function to model the background due to natural radioactivity , and a Polya function, expressed by Eq.~\ref{fun:polya}, often used to describe the response spectrum of MPGD \cite{bib:rolandiblum}:
\begin{equation}
   P(n)=\frac{1}{b\overline{n}}\frac{1}{k!}\left(\frac{n}{b\overline{n}}\right)^k \cdot e^{-n/b\overline{n}}
   %A\frac{\Gamma(r+k)}{k!\Gamma(r)} p^k (1-p)^r e^{(-x/\lambda)} 
\label{fun:polya}
\end{equation}
%
%where n represents the number of secondary electrons, 
where $b$ is a free parameter and $k=1/b-1$. The distribution has $\overline{n}$ as expected value, while the variance is governed by its mean and the parameter $b$: $\sigma^2=\overline{n}(1+b\overline{n})$. 
From the result of the fits it is possible to evaluate:
\begin{itemize}
\item the expected value of the distribution $\overline{n}$ and its variance $\sigma^2$. These parameters, when fitted on the light distributions, give the detector response in terms of number of photons and the energy resolution.  
When fitting the number of illuminated pixels distribution, the average size of the clusters can be evaluated by taking into account the effective area of $130~\times 130~\mu$m$^2$ seen by each single pixel.
\item the integral of the Polya component, that is proportional to the total number of reconstructed clusters and that can be used to evaluate the detection efficiency;
\end{itemize}
Since, as it is shown on the left of Fig.~\ref{fig:spectra}, in this configuration 1169~$\pm$~223 photons are collected per cluster (i.e. each 5.9~keV released), a threshold of 400 photons corresponds to about 2~keV released in the sensitive volume.
The average cluster size was found to be 149 pixels (Fig.~\ref{fig:spectra}, right). 

\subsection{Sensor electronic noise}
The CMOS sensor used for the measurements has two main sources of noise: 
\begin{itemize}
    \item a dark current of about 0.06 electrons per second per pixel;
    \item a readout noise of about 1.4 electrons rms (in our set-up it was found to be slightly
    larger probably due to an effect of {\it ageing} of the sensor built more than 5 years ago);
\end{itemize}

\begin{figure}[htbp]
\centering
\includegraphics[width=.45\textwidth]{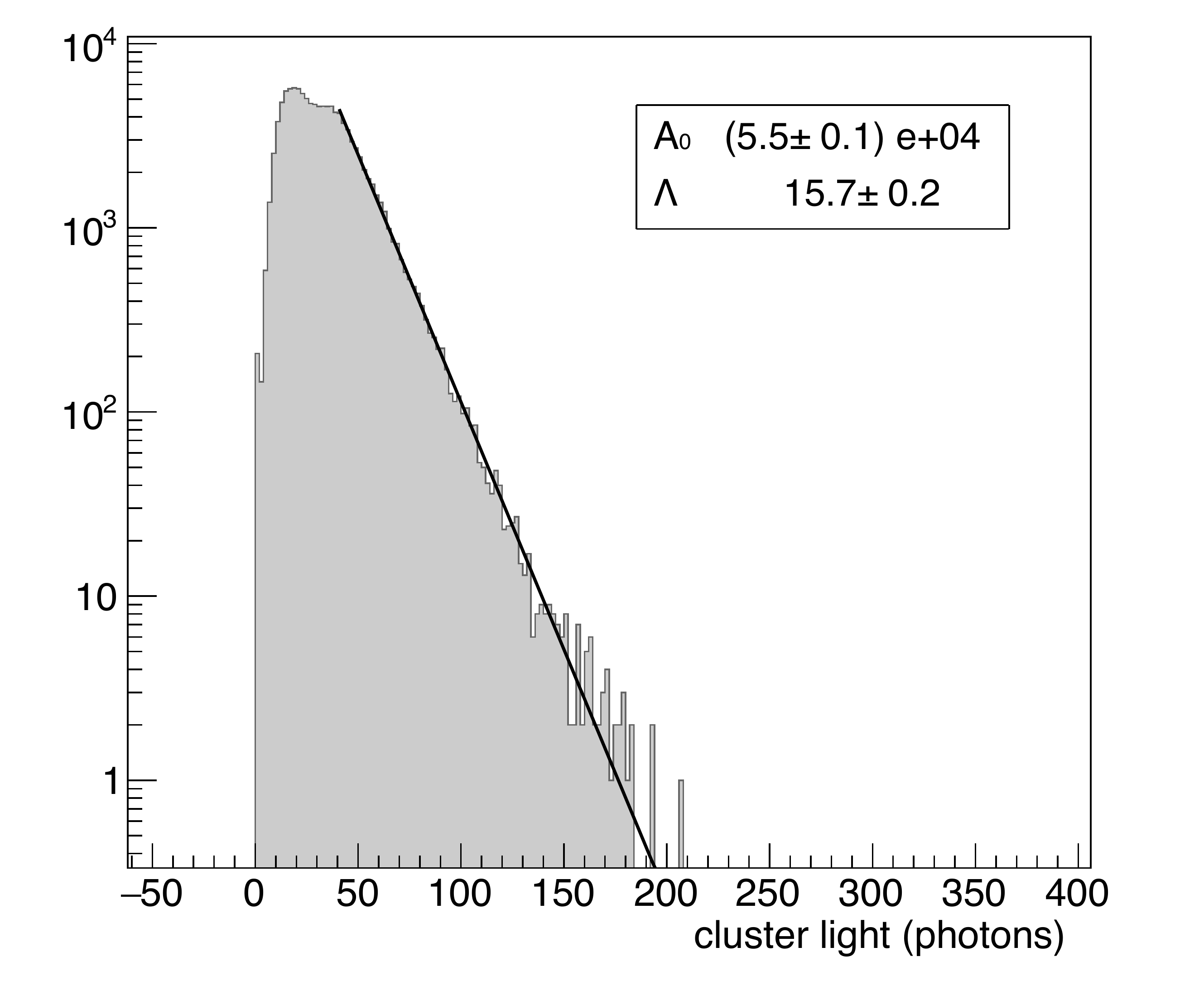}
\caption{Distribution of the light in clusters reconstructed in a run with blind sensor.}
\label{fig:hq_ghost}
\end{figure}

The sensor electronic noise represents a possible
unavoidable instrumental background and it can generate {\it ghost-clusters}.
%In Fig.\ref{fig:map_ghost}, the map of ghost-clusters reconstructed.
The distribution of the light in each {\it ghost-cluster} found in the {\it blind run} 
is shown in Fig.~\ref{fig:hq_ghost}. 
It shows a baseline component due to positively-definite counts of photons and to the minimal number of two macro-pixels requested to form a cluster. In order to define an operative threshold that allows to suppress fake signals due to sensor noise, the tail of this distribution is studied. In particular, it is fitted with an exponential function:
\begin{equation}
    p(L)=A_0~e^{-\frac{L}{\Lambda}}
\label{eq:exp}
\end{equation}
where $L$ is the number of photons collected in the cluster. 

From the fitted parameters, $A_0~=~(5.5~\pm~0.1)~\times~10^4$ and $\Lambda=15.7~\pm~0.4$~photons, and by taking into account that a run lasts 10 s, it is possible to extrapolate the probability of having a {\it ghost-clusters} with an amount of light larger than a given threshold. Results for three threshold values are shown in Table~\ref{tab:thr_1}.

\begin{table}[htbp]
\centering
\caption{Number of {\it ghost-clusters} found per sensor per second containing a total amount of light over a certain threshold as extrapolated from the fit in Fig.\ref{fig:hq_ghost}.}
\vspace{0.2 cm}
\begin{tabular}{ |c|c| } 
 \hline
 Threshold (photons) & {\it ghost-clusters}/second \\ 
 \hline
 \hline
 200 & 2$\times$10$^{-2}$  \\ 
 300 & 1$\times$10$^{-4}$  \\ 
 400 & 3$\times$10$^{-7}$  \\ 
 \hline
\end{tabular}
\label{tab:thr_1}
\end{table}

More detailed analysis (e.g. topological light distribution) can be further exploited to discriminate clusters induced by signals from {\it ghost-clusters}. This information has not been used in the data analysis presented in this paper, but it will provide additional handle to suppress the backgrounds in the future.

\section{Combined Optical Read-out}

To overcome the poor timing information provided by the CMOS,
a combined light readout with a fast PMT was tested.

\begin{figure}
\centering
\includegraphics[width=.75\textwidth]{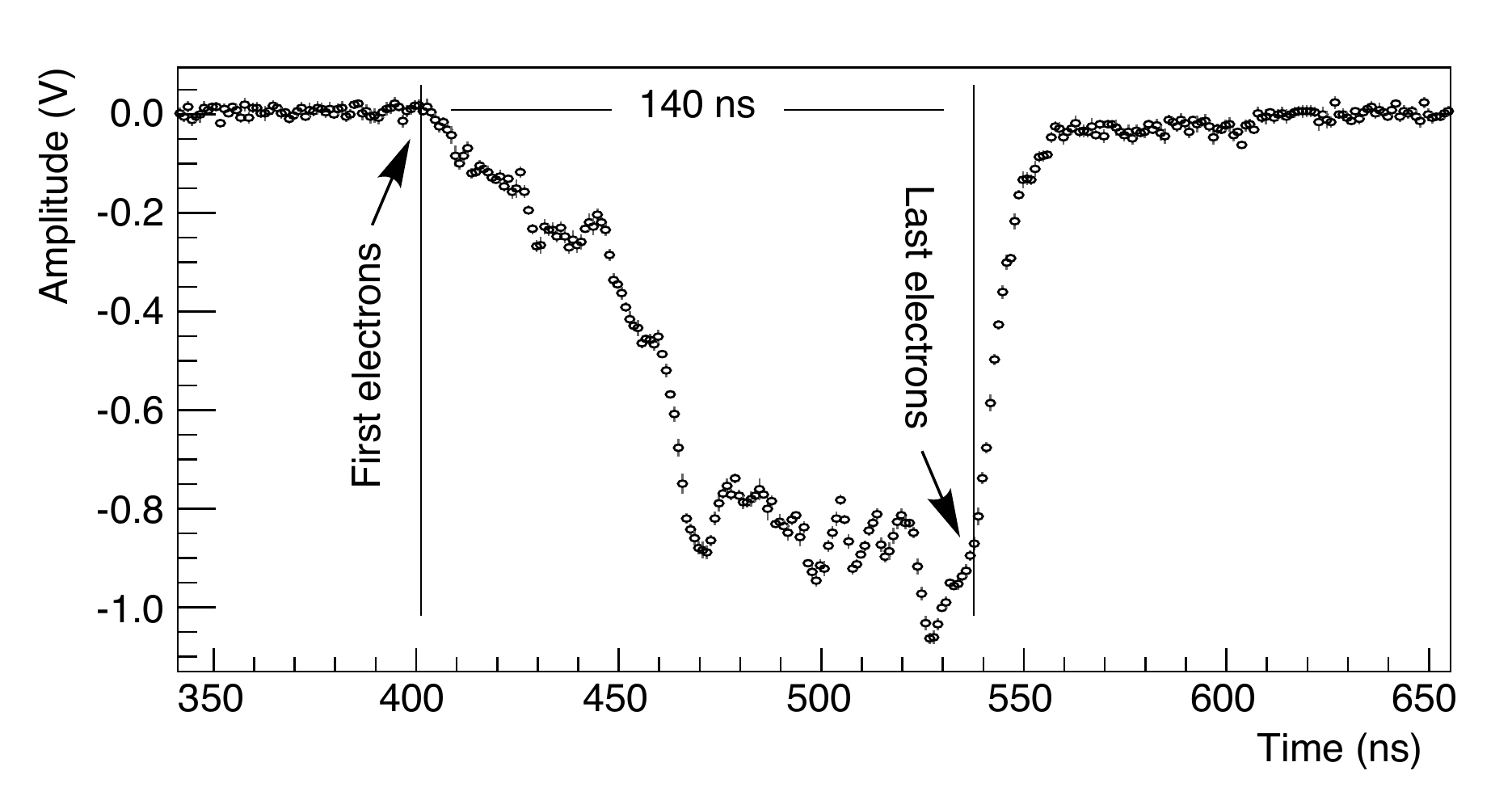}
\caption{PMT waveform 
for a track crossing the drift gap inclined with respect to the GEM plane.}
\label{fig:pmt-inclined}
\end{figure}

In Fig. \ref{fig:pmt-inclined}, the PMT signal 
is shown for an inclined electron crossing the 1 cm drift gap
at an angle of 0.1 rad (almost 6$^{\circ}$) with respect to the GEM foils.

The arrival time of the main clusters is clearly visible,
allowing an independent reconstruction of their 
absolute position in {\it z}.
Taking into account the gap width (1 cm)
and the width of the signal (about 135 ns),
an electron drift velocity of 72 $\mu$m/ns
is found in agreement with the value 
evaluated with Garfield.

\begin{figure}
\centering
\includegraphics[width=.45\textwidth, angle=90]{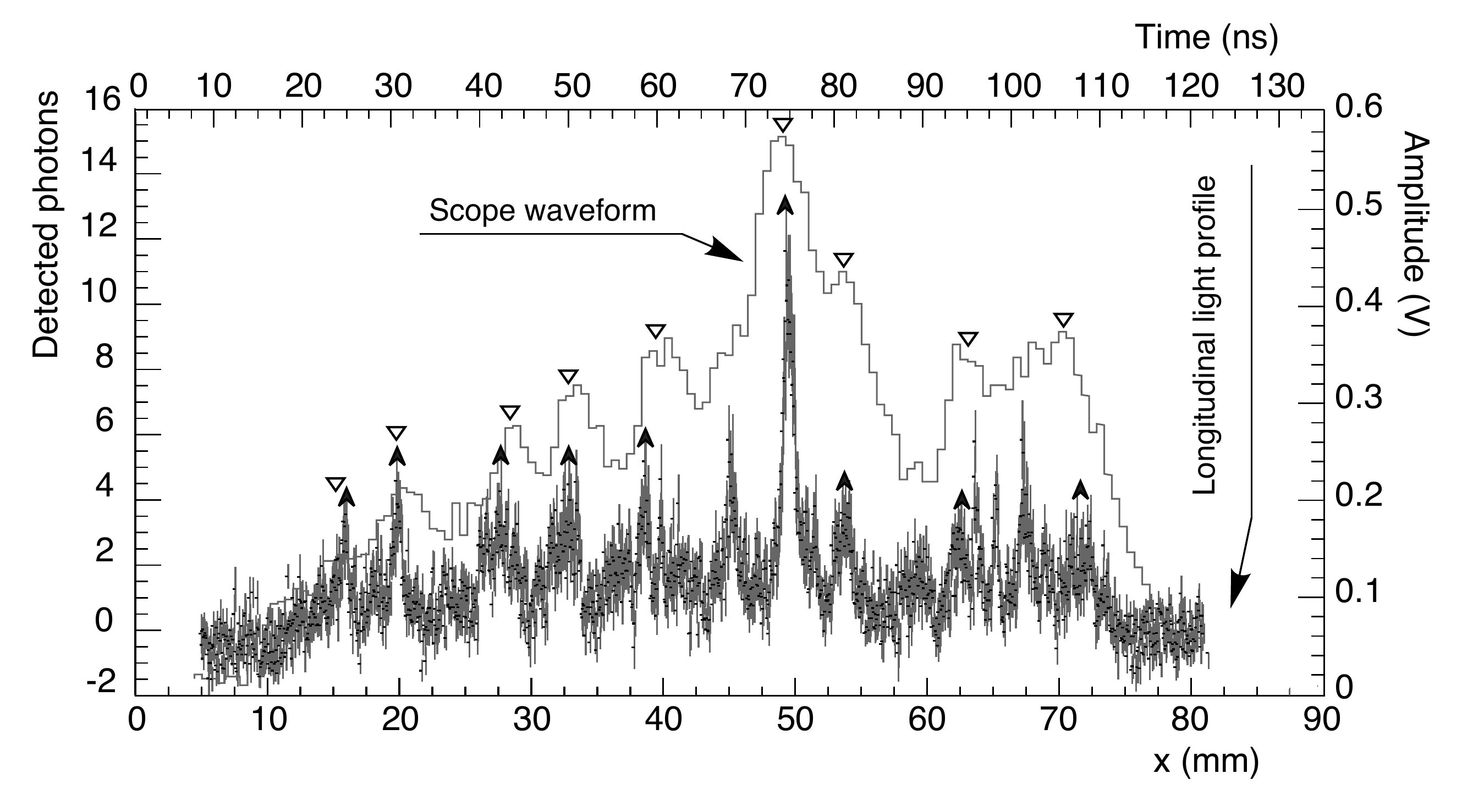}
\caption{Lateral profile of the light detected by the CMOS sensor along with 
the waveform of the PMT signal for the same event.
The cluster structure is clearly visible in both cases.
Peaks found by the algorithm are shown.}
\label{fig:peak}
\end{figure}

Figure \ref{fig:peak} shows an example of 
the lateral profile of the detected light 
as seen by the CMOS camera for a track
together with the corresponding PMT waveform.

In both cases, with a simple peak finding algorithm, 
the position of the main peaks was evaluated. 
The tracks in the ten analysed events 
have an average length of almost 60 mm
and 54 peaks were found in total. Therefore, the algorithm is
able to identify one peak per track centimetre on average.

\begin{figure}
\centering
\includegraphics[width=.85\textwidth]{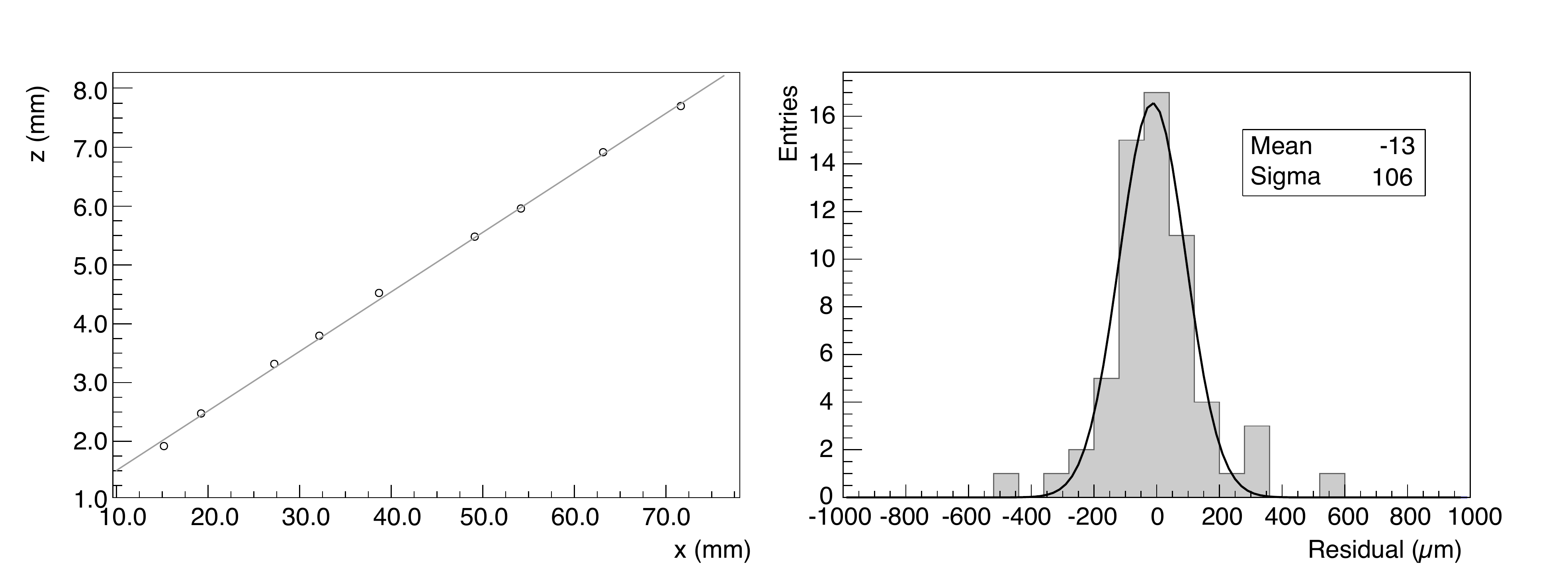}
\caption{
Left: Correlation with a superimposed linear fit of the
{\it x} and {\it z} coordinates of the clusters found in a single track.
Right: Distribution of the residuals of the reconstructed to the linear fit
{\it z} for ten tracks with a superimposed gaussian fit.}
\label{fig:zres}
\end{figure}

By assuming these peaks as due to ionization 
clusters along the tracks,
their {\it x} and {\it z} coordinates can be evaluated.
Their correlation is shown on the left of Fig. \ref{fig:zres}
while the distribution of the reconstructed
{\it z} residuals to a linear fit for a set of ten tracks 
(used as an example) is shown on the right.

From the superimposed fit, it is possible to evaluate a resolution 
on the reconstructed {\it z} coordinate of about 100 $\mu$m.

\section{Operation Stability}

The stability of the operation of the prototype was tested in laboratory.
High voltage values in the detector were set to working conditions. For 13 days the detector was kept operative while the currents and environmental parameters were continuously monitored by an automated system and images due to natural radioactivity were recorded.
From time to time, some hot-spots, due to micro-discharges in imperfect channel, appear on the GEM surface. Usually they do not involve large currents, but they can grow up making the detector drawing a sizable current. An automatic “recovery” procedure was implemented able to turn OFF these hot-spots by lowering and gradually restoring the HV conditions in few minutes. Together with these slow phenomena, the detector had some sudden jump in current rapidly reset due to discharge.

\begin{figure}[htbp]
\centering
\includegraphics[width=.65\textwidth]{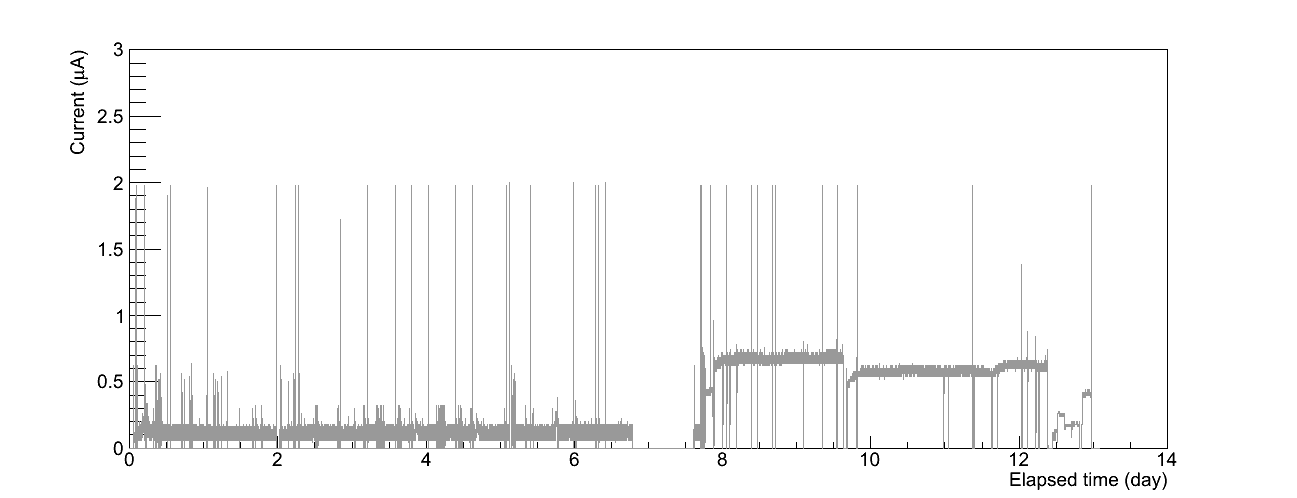}
\caption{Plot of the currents drawn by the third GEM (see text for details).}
\label{fig:stab}
\end{figure}

Figure \ref{fig:stab} shows an example of the behavior of the detector current.

In total 28 discharges (3/day) and 72 hot spots (8/day) are visible during the acquisition period.
Apart from these events, the current does not show any instabilities and detector operation was rather stable. The recovery procedure lasted for a total of 15 hours making the detector live-time 94\%.

\section{The CYGNO project}

\begin{figure}[htbp]
\centering
\includegraphics[width=.65\textwidth]{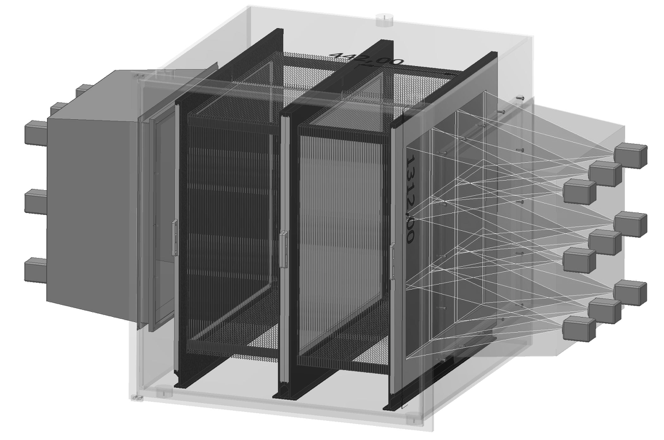}
\caption{Drawing of the 1 cubic meter {\it CYGNO} apparatus.}
\label{fig:cygno}
\end{figure}

This technology showed to be really promising to develop a detector for Directional Light Dark Matter search.
The drafting of a Technical Design Report was funded to describe a 3/4 year project leading to construction of CYGNO, a 1 m$^3$ TPC based on optical readout 
(shown in \ref{fig:cygno}).

One m$^3$ of He/CF$_4$ 60/40 (1.6 kg) at atmospheric pressure subdivided in two 50 cm long parts by the cathode with a drift field of about 1 kV/cm.
Each gas volume is equipped by a 3$\times$3 matrix with triple-GEM structure readout by: 
\begin{itemize}
    \item a sCMOS sensor 65 cm away from a transparent window;
    \item a fast light detector (PMT or SiPM).
\end{itemize}

The active apparatus will be contained in shields for gamma ray and neutrons.

CYGNO will behave as a demonstrator of the technology performance in order to prepare a proposal  for a 30-100 m$^3$ experiment for Directional Dark Matter search in the low mass region that can represent the first module of a World Wide Network of underground observatories in the CYGNUS-TPC framework \cite{bib:0}.

\section*{References}

\bibliographystyle{iopart-num}
\bibliography{fe55}{}

%\section*{References}
%\begin{thebibliography}{9}
%\bibitem{iopartnum} IOP Publishing is to grateful Mark A Caprio, Center for %Theoretical Physics, Yale University, for permission to include the {\tt %iopart-num} \BibTeX package (version 2.0, December 21, 2006) with  this %documentation. Updates and new releases of {\tt iopart-num} can be found on %\verb"www.ctan.org" (CTAN). 
%\end{thebibliography}

\end{document}